\documentclass[aps,prl,reprint,superscriptaddress]{revtex4-1}

\usepackage{graphicx}
\usepackage{textcomp}
\usepackage{color}

\begin{document}

\title{
Highly coherent electron beam from a laser-triggered tungsten needle tip}

\author{Dominik Ehberger}
\altaffiliation{D.E. and J.H. contributed equally to this work.}
\altaffiliation{now with Ludwig-Maximilians-Universit{\"a}t M{\"u}nchen, Am Coulombwall 1, 85748 Garching, Germany, EU}
\affiliation{Department of Physics, Friedrich Alexander University Erlangen-Nuremberg, Staudtstr. 1, D-91508 Erlangen, Germany, EU}
\affiliation{Max Planck Institute for Quantum Optics, Hans-Kopfermann-Str. 1, D-85748 Garching/Munich, Germany, EU}

\author{Jakob Hammer}
\altaffiliation{D.E. and J.H. contributed equally to this work.}
\affiliation{Department of Physics, Friedrich Alexander University Erlangen-Nuremberg, Staudtstr. 1, D-91508 Erlangen, Germany, EU}
\affiliation{Max Planck Institute for Quantum Optics, Hans-Kopfermann-Str. 1, D-85748 Garching/Munich, Germany, EU}

\author{Max Eisele}
\altaffiliation{now with Department of Physics, University of Regensburg, 93040 Regensburg, Germany, EU}
\affiliation{Max Planck Institute for Quantum Optics, Hans-Kopfermann-Str. 1, D-85748 Garching/Munich, Germany, EU}

\author{Michael Kr{\"u}ger}
\altaffiliation{now with Department of Physics of Complex Systems, Weizmann Institute of Science, Rehovot 76100, Israel}
\affiliation{Department of Physics, Friedrich Alexander University Erlangen-Nuremberg, Staudtstr. 1, D-91508 Erlangen, Germany, EU}
\affiliation{Max Planck Institute for Quantum Optics, Hans-Kopfermann-Str. 1, D-85748 Garching/Munich, Germany, EU}

\author{Jonathan Noe}
\affiliation{Fakult{\"a}t f{\"u}r Physik and Center for NanoScience (CeNS), Ludwig-Maximilians-Universit{\"a}t M{\"u}nchen, Geschwister-Scholl-Platz 1, 80539 M{\"u}nchen, Germany, EU}

\author{Alexander H{\"o}gele}
\affiliation{Fakult{\"a}t f{\"u}r Physik and Center for NanoScience (CeNS), Ludwig-Maximilians-Universit{\"a}t M{\"u}nchen, Geschwister-Scholl-Platz 1, 80539 M{\"u}nchen, Germany, EU}

\author{Peter Hommelhoff}
\affiliation{Department of Physics, Friedrich Alexander University Erlangen-Nuremberg,  Staudtstr. 1, D-91508 Erlangen, Germany, EU}
\affiliation{Max Planck Institute for Quantum Optics, Hans-Kopfermann-Str. 1, D-85748 Garching/Munich, Germany, EU}
\affiliation{Max Planck Institute for the Science of Light, G\"unther-Scharowsky-Str. 1/ Bldg. 24, D-91508 Erlangen, Germany, EU}

\date{\today}

\begin{abstract}

We report on a quantitative measurement of the spatial coherence of electrons emitted from a sharp metal needle tip. We investigate the coherence in photoemission using near-ultraviolet laser triggering with a photon energy of $3.1\,$eV  and compare it to DC-field emission. A carbon-nanotube is brought in close proximity to the emitter tip to act as an electrostatic
biprism. From the resulting electron matter wave interference fringes we
deduce an upper limit of the effective source radius both in
laser-triggered and DC-field emission mode, which quantifies the spatial
coherence of the emitted electron beam. We obtain $(0.80\pm0.05)\,$nm in
laser-triggered and $(0.55\pm0.02)\,$nm in DC-field emission mode,
revealing that the outstanding coherence properties of electron beams from
needle tip field emitters are largely maintained in laser-induced emission.
In addition, the relative coherence width of 0.36 of the photoemitted
electron beam is the largest observed so far. The preservation of
electronic coherence during emission as well as ramifications for
time-resolved electron imaging techniques are discussed.

\end{abstract}


\maketitle

Coherent electron sources are central to studying microscopic objects with highest spatial resolution. They provide electron beams with flat wavefronts that can be focused to the fundamental physical limit given by matter wave diffraction~\cite{Spence2013}. Currently, \textit{time-resolved} electron based imaging is pursued with large efforts, both in real-space microscopy~\cite{Zewail2010a, King2005} and in diffraction~\cite{Sciaini2011, Baum2007b}. However, the spatial resolution in time-resolved electron microscopy is about two orders of magnitude worse than its DC counterpart~\cite{Yurtsever2012}, which reaches below ${0.1}\,$\AA~\cite{Erni2009}. Combining highest spatial resolution with time resolution in the picosecond to (sub-) femtosecond range requires spatially coherent electron sources driven by ultrashort laser pulses. Although laser-driven metal nanotips promise to provide coherent electron pulses with highest time resolution, a quantitative study of their spatial coherence has been elusive. 
Here we demonstrate that photoemitted electrons from a tungsten nanotip are highly coherent.

So far no time-resolved electron based imaging instrument fully utilizes the coherence capabilities provided by nanotip electron sources. Meanwhile, nanotips operated in DC-field emission are known and employed in practical applications for almost half a century for their paramount spatial coherence properties~\cite{Crewe1968}. Thence, highest resolution microscopy as well as coherent imaging, such as holography and interferometry, have long been demonstrated in DC-field emission~\cite{Spence2013, Lichte2008, Hasselbach2010}. Here we investigate whether these concepts can be inherited to laser-driven nanotip sources by comparing the spatial coherence of photoemitted electron beams to their DC counterparts. This would enable time-resolved high resolution imaging, but may also herald fundamental studies based on the generation of quantum degenerate electron beams~\cite{Lougovski2011}.

\begin{figure}
\includegraphics[width=86mm]{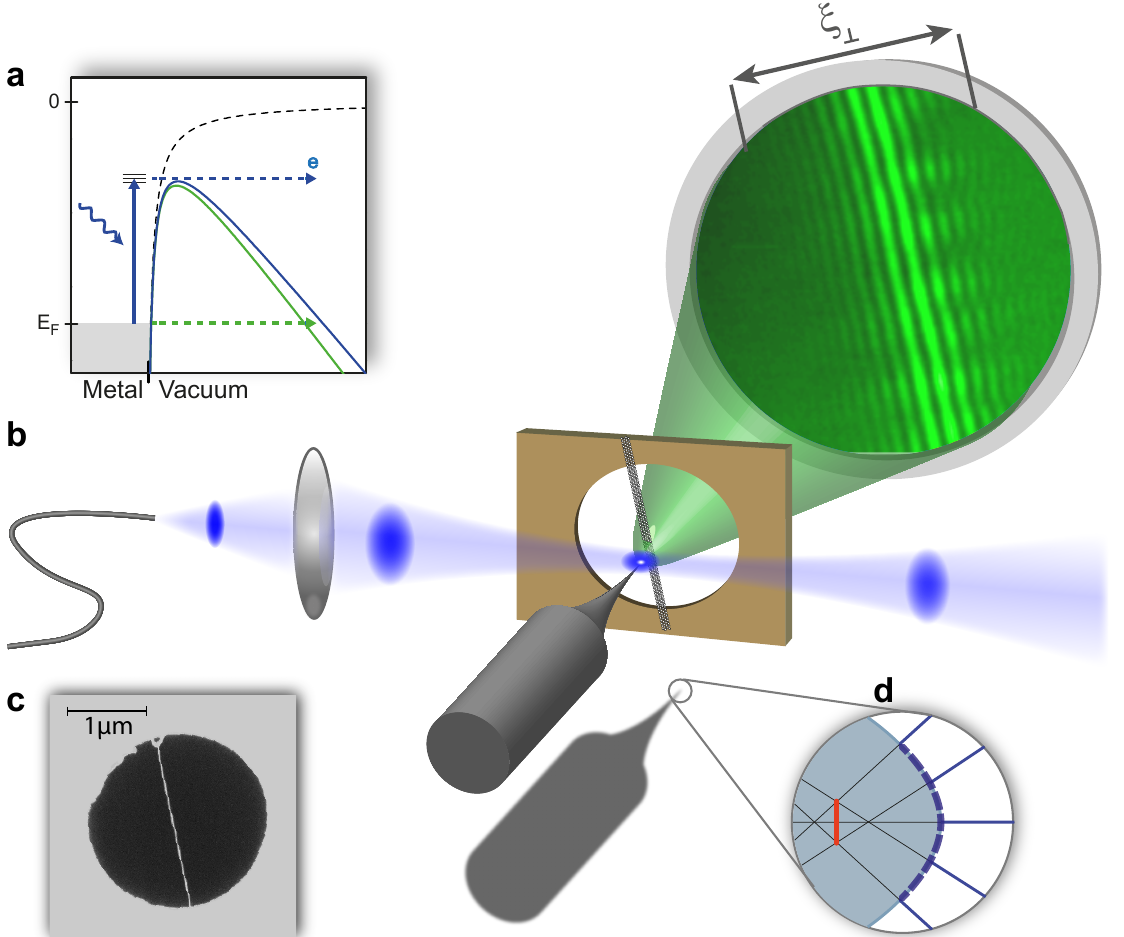}
\caption{(Color online) Schematic of the experimental setup. (\textbf{a}) 
Illustration of one-photon photoemission (blue) and DC-field emission (green). Electrons from states around the Fermi-level $E_\mathrm{F}$ are excited by laser irradiation and emitted over the barrier, which is lowered due to the Schottky effect. At sufficiently high DC-fields the barrier becomes narrow enough to permit direct tunneling through it, giving rise to DC-field emission. (\textbf{b}) Ultrahigh vacuum setup. A near-UV laser beam (photon energy of $3.1\,$eV) coupled into the chamber via a polarization maintaining fiber is focused onto the apex of a  tungsten tip. Interference patterns are obtained on the microchannel plate detector after the electrons have passed a freestanding carbon nanotube (CNT) beam splitter placed in close vicinity to the tip.  (\textbf{c}) Scanning electron microscope image of a CNT grown over a hole of a SiN membrane. (\textbf{d}) A virtual (or effective) source (vertical red full line) is formed behind the tip's apex, substantially smaller than the geometrical source size (blue dashed line). Solid lines indicate electron trajectories.}
\label{setup}
\end{figure}

The spatial coherence of electron sources is commonly quantified by means of their effective source radius $r_\mathrm{eff}$.
It equals the radius of a virtual incoherent emitter that resembles the coherence properties of the real emitter. As discussed later, $r_\mathrm{eff}$ is inversely proportional to the transverse coherence length $\xi_\perp$ of the electron beam. 
A virtual source is formed in a finite area where electron trajectories intersect when extrapolating their paths back into the metal tip (Fig.~\ref{setup}d). For tungsten field emitters
typical values for $r_\mathrm{eff}$ are on the order of $1\,$nm and the smallest reported down to $0.4\,$nm, significantly smaller than the geometrical tip radius that is typically in the range of a few tens of nanometers~\cite{Spence2013, Cho2004}. 

DC-field and laser-driven emission occur due to fundamentally different emission processes (Fig.~\ref{setup}a)~\cite{Jensen2010}. The former is a tunnelling process through a static potential barrier, covered within the Fowler-Nordheim-theory~\cite{Fowler1928}, whereas a variety of laser-driven emission processes exist. They are distinguishable into linear one-photon emission and nonlinear multi-photon and tunneling processes, with the respective prominent examples of Einstein's photoelectric effect and multi-photon emission~\cite{Delone1994}. 
The effective source radius is highly sensitive to
the shape of the electron trajectories in close vicinity of the tip apex~\cite{Cook2010}, and hence to the emission process. As a result, the coherence properties in
photoemission might be drastically different from DC-field emission.

To compare the coherence properties of a monocrystalline tungsten tip electron emitter with a radius of $\sim$10\,nm in laser-triggered and DC-field emission we record electron matter wave interference images in both emission modes. We use a freestanding carbon nanotube (CNT) as an electron beam splitter, which acts as a biprism filament with nanometer radius~\cite{Hasselbach2010}. It splits the wavefront of the electron matter wave in two parts, which are then overlapped at the electron detector, giving rise to interference fringes on the detector screen~\cite{Cho2004}. 
A scanning electron microscope image of a single, freestanding CNT on a holey silicon nitride membrane is shown in Fig.~\ref{setup}c (see Supplementary Material for details). The electrically grounded CNT is brought into the electron beam path at a typical distance of less than one micrometer from the tip, resembling a point projection microscopy configuration, that is also commonly used for electron holography~\cite{Beyer2010,Longchamp2013}. CNT and the gold coated holey silicon nitride membrane act as a counter electrode for the biased tip.  Electron interference can be observed in conventional DC-field emission as well as in laser-triggered mode when a near-UV laser beam is focused on the tip's apex (see Fig.~\ref{setup}b and Supplementary Material for details).

\begin{figure}
\includegraphics[width=86mm]{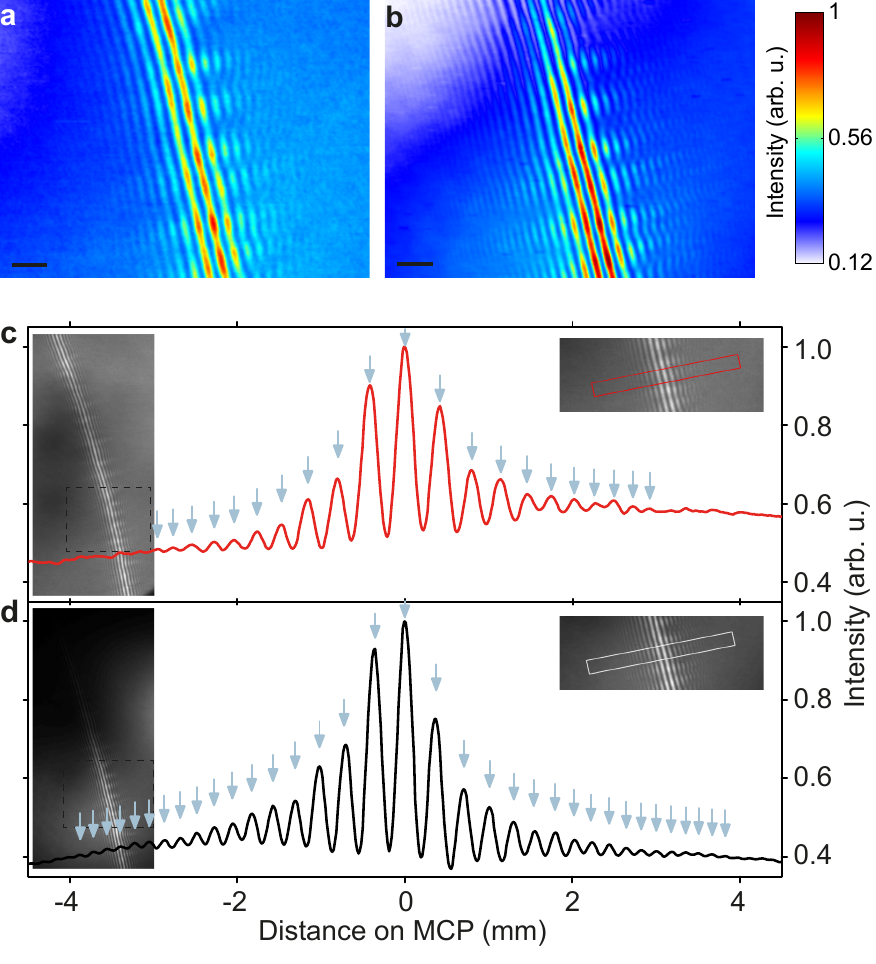}
\caption{(Color online) Electron interference fringes on the detector screen. 
(\textbf{a}) Laser-triggered electron emission at a bias voltage of $U_\mathrm{tip}=-41\,$V. Without laser illumination no electrons are observed at this voltage. Scale bar is $1\,$mm on the detector screen. (\textbf{b}) DC-field emission (  $U_\mathrm{tip}=-53\,$V). Images are obtained by superimposing 200 individual images that have been corrected for slow linear drifts (see Supplementary Material). A modulation of the fringe pattern along the CNT direction is also clearly discernible, arising from local distortions of the CNT and locally enhanced DC-fields. 
Line profiles of the interference fringes are integrated perpendicular to the orientation of the CNT with laser-triggered (\textbf{c}) and DC-field emission source  (\textbf{d}). The box in the right inset indicates the 9.3\,mm$^2$ large integration area. The left inset shows a larger detector image. At least 21 fringes in the laser-triggered mode and 35 in DC-field emission mode are visible as indicated with arrows. The slightly finer spacing in DC-field emission is due to the smaller electron de Broglie wavelength (see Supplementary Material).}
\label{diff_im}
\end{figure}

An upper bound for $r_\mathrm{eff}$ is obtained by measuring the full width $\xi_\perp$ of coherent illumination at the detector screen. It can be deduced by identifying the distance between the outermost interference fringes, observed perpendicular to the orientation of the CNT~\cite{Spence1994}. 
The van Cittert-Zernicke theorem relates $\xi_\perp$ and the effective source radius $r_\mathrm{eff}$ for an incoherent emitter with Gaussian intensity profile~\cite{Pozzi1987,Spence1994}:
\begin{equation}
r_\mathrm{eff}= \frac{\lambda_\mathrm{dB}\cdot l_\mathrm{s-d}}{\pi\cdot\xi_\perp}.
\end{equation}
Here $\lambda_\mathrm{dB}$ is the electron de Broglie wavelength and $l_\mathrm{s-d}$ the source-detector distance.

Ideally, an electron source should exhibit a narrow (longitudinal) momentum distribution to reduce chromatic effects in subsequent electron optics. Hence, in order to achieve efficient electron emission with little excess momentum we match the Schottky-lowered barrier at the metal-vacuum interface, which is tunable by means of the tip voltage, to the photon energy ($E_\mathrm{ph}=3.1\,$eV) of the focused laser light. Here the barrier height is set to $2.8\pm0.1\,$eV, closely above the onset of DC-field emission and yielding the highest photoemission probability with negligible DC component. The experiments are performed with a pulsed laser (second harmonic of $130\,$fs long pulses derived from a 2.7\,MHz repetition rate long-cavity Ti:sapphire oscillator) and a high power cw-diode laser source ($400\,$mW at 405\,nm) to boost the electron current for the effective source radius measurements.

Electron interference patterns in laser-triggered and DC-field emission are shown in Fig.~\ref{diff_im}a and b, respectively, recorded at the identical CNT position with a tip-sample distance of less than $1\,$micrometer. Clearly, interference fringes that are aligned parallel to the CNT are observed in both modes. A tip voltage of $U_{\mathrm{tip}}=-41\,$V is chosen in laser-induced emission, such that the barrier is lowered for efficient photoemission. For DC-field emission a voltage of $U_{\mathrm{tip}}=-53\,$V is applied, leading to a comparable field emission electron current as in photoemission.

The panels in Fig.~\ref{diff_im}c,\,d show line profiles obtained from integrating the count rate parallel to the fringes in the marked area.
The spatial coherence width is obtained from these line profiles. For photoemission we obtain
$\xi_\perp^\mathrm{ph}\geq 5.9\,$mm
at a CNT-screen distance of $79.5\,$mm. With 
$\lambda_\mathrm{dB} = 1.8\,$\AA, the effective source radius equals $r_\mathrm{eff}^\mathrm{ph}\leq 0.80\pm0.05\,$nm.

In DC-field emission mode a very comparable value
of the coherence width is deduced with
$\xi_\perp\geq 7.7\,$mm,
 albeit slightly larger (Fig.~\ref{diff_im}d). 
With $\lambda_\mathrm{dB}=1.7\,$\AA\ the effective source radius equals $r_\mathrm{eff}^\mathrm{DC}\leq 0.55\pm0.02\,$nm, in line with previously published values~\cite{Spence1994,Cho2004}.

Clearly, the source radii in laser-triggered and DC-field emission mode differ only slightly, even though the emission process is qualitatively different.
Furthermore, in both cases the effective source radius is about an order of magnitude smaller than the geometrical source radius.
For comparison, the record resolution laser-triggered electron microscope employs a fully illuminated flat $\mathrm{LaB}_6$ cathode of a few tens of microns in diameter~\cite{Barwick2008, NoteCommercialLasertriggeredSEM}. In this configuration the effective source radius equals the geometric one, given by the smaller of either the laser spot size or the dimensions of the cathode.

The  relative coherence width $K$, namely the ratio of $\xi_\perp$ to the electron beam radius $R_\mathrm{I}$, is a conserved quantity in electron optics~\cite{Pozzi1987}. Thus, it allows calculating $\xi_\perp$ for any given beam size, in particular for arbitrary focusing conditions at a sample. With $R_\mathrm{I}=16\,$mm ($1/e^2$-radius of the electron beam) at the detector the relative coherence width of the photoemitted beam equals $0.36$, representing the highest value reported for $K$ of a laser-triggered electron source to date. It benefits largely from the use of a monocrystalline tip, which exhibits a low divergence of the emitted beam of $\sim\,10^\circ$ (half angle) in photoemission and $\sim\,6^\circ$ in DC-field emission.

With increasing electron current it can be expected that the effective source size increases due to space charge and stochastic Coulomb electron-electron repulsion~\cite{Cook2010}. Strictly, these effects come into play for more than one electron per pulse emitted from the tip. Hence most conservatively, the maximum current attainable with highest spatial coherence is set by the repetition rate $f_\mathrm{rep}$ of the laser. For instance, laser pulses with $f_\mathrm{rep}=100\,$MHz inducing emission of one electron per pulse yield a time averaged current of $16\,$pA. Even though this value is low compared to the electron current emitted from standard field emission guns 
electron imaging with a stably aligned laser beam, as demonstrated here, remains well possible as demonstrated in time-resolved scanning electron microscopy
~\cite{Yang2010}. The restriction to one electron per pulse, however, also prevents other unwanted detrimental effects such as temporal electron pulse broadening due to Coulomb repulsion~\cite{Kirchner2014}. 
In this experiment, the required minimum peak fluence $2 E_\mathrm{P}/(\pi w_0^2)$ to obtain one electron per pulse without DC contributions equals $0.2\,\mathrm{J}/\mathrm{cm}^2$, with the pulse energy $E_\mathrm{P}$ and the $1/\mathrm{e}^2$-beam waist radius $w_0$. Note that many more than one electron per pulse can be drawn from the tip for most settings without detrimental effects on the beam quality, especially after propagation to a sample. This, however, depends on various parameters such as tip radius, laser pulse duration, acceleration field and  electron beam path.

Next to the transverse coherence, quantified by $r_\mathrm{eff}$, the energy spread of the electron beam $\Delta E$ is crucially important for most applications. 
Here $\Delta E$ of the photoemitted beam equals $0.51\pm0.06\,$eV (FWHM), less than twice as much as in DC-field emission~\cite{Spence2013}. This implies that the longitudinal coherence length  is smaller by a factor of about 2 in photoemission~\cite{Lichte2008}, likely causing the reduced visibility of the interference pattern in Fig.~\ref{diff_im}c (see Supplementary Material). 
We find that in principle the energy spread can be made as low as in DC-field emission with constant electron current by decreasing the DC-field at the tip and simultaneously increasing the laser power (see Supplementary Material). For instance, here $\Delta E\approx 0.4\,$eV is feasible with an increased barrier height of $\sim 3.0\,$eV and tripled laser fluence.

We conclude that the coherence of the electron beam in one-photon photoemission close to the threshold is almost as good as that of a DC-field emitted beam. It has been previously shown that the initial electronic states inside the metal from which the electrons originate affect the coherence of the emitted electron beam~\cite{Cho2004}. Our measurements demonstrate that the coherence of the original electronic states inside the metal is maintained in photoemission. One may thus expect that a cooled tip also provides a fully coherent beam under laser irradiation, as demonstrated in DC-field emission~\cite{Cho2004}.

By virtue of the excellent coherence properties of a DC-field emitted electron beam it was shown that the combination of point projection holography and coherent electron diffraction allows for $2\,$\AA~resolution in imaging of graphene~\cite{Longchamp2013}. Very recently, first time-resolved results have been obtained in femtosecond point projection microscopy~\cite{Quinonez2013}, ultrafast low-energy electron diffraction~\cite{Gulde2014} and combinations of both~\cite{Muller2014} based on femtosecond laser-triggered tungsten field emission tips as electron sources~\cite{Hommelhoff2006,Hommelhoff2006a,Ropers2007}. In this context our findings clearly show that electron imaging devices equipped with field emission guns can be laser-triggered to obtain highest temporal resolution without losing their supreme coherence and imaging properties. 
The excellent source properties will also be of great interest for novel laser-based electron acceleration schemes as recently demonstrated~\cite{Peralta2013,Breuer2013}.

The authors thank H. Kaupp for discussions on electron beam coherence measurements prior to the experiment,  S. Stapfner, L. Ost and E. Weig for discussions on CNT fabrication and  J.P. Kotthaus for cleanroom access.
This research is funded in part by the the Gordon and Betty Moore Foundation, the DFG Cluster of Excellence Munich Centre for Advanced Photonics and the ERC Grants NearFieldAtto and QuantumCANDI.


\begin{thebibliography}{32}%
\makeatletter
\providecommand \@ifxundefined [1]{%
 \@ifx{#1\undefined}
}%
\providecommand \@ifnum [1]{%
 \ifnum #1\expandafter \@firstoftwo
 \else \expandafter \@secondoftwo
 \fi
}%
\providecommand \@ifx [1]{%
 \ifx #1\expandafter \@firstoftwo
 \else \expandafter \@secondoftwo
 \fi
}%
\providecommand \natexlab [1]{#1}%
\providecommand \enquote  [1]{``#1''}%
\providecommand \bibnamefont  [1]{#1}%
\providecommand \bibfnamefont [1]{#1}%
\providecommand \citenamefont [1]{#1}%
\providecommand \href@noop [0]{\@secondoftwo}%
\providecommand \href [0]{\begingroup \@sanitize@url \@href}%
\providecommand \@href[1]{\@@startlink{#1}\@@href}%
\providecommand \@@href[1]{\endgroup#1\@@endlink}%
\providecommand \@sanitize@url [0]{\catcode `\\12\catcode `\$12\catcode
  `\&12\catcode `\#12\catcode `\^12\catcode `\_12\catcode `\%12\relax}%
\providecommand \@@startlink[1]{}%
\providecommand \@@endlink[0]{}%
\providecommand \url  [0]{\begingroup\@sanitize@url \@url }%
\providecommand \@url [1]{\endgroup\@href {#1}{\urlprefix }}%
\providecommand \urlprefix  [0]{URL }%
\providecommand \Eprint [0]{\href }%
\providecommand \doibase [0]{http://dx.doi.org/}%
\providecommand \selectlanguage [0]{\@gobble}%
\providecommand \bibinfo  [0]{\@secondoftwo}%
\providecommand \bibfield  [0]{\@secondoftwo}%
\providecommand \translation [1]{[#1]}%
\providecommand \BibitemOpen [0]{}%
\providecommand \bibitemStop [0]{}%
\providecommand \bibitemNoStop [0]{.\EOS\space}%
\providecommand \EOS [0]{\spacefactor3000\relax}%
\providecommand \BibitemShut  [1]{\csname bibitem#1\endcsname}%
\let\auto@bib@innerbib\@empty
\bibitem [{\citenamefont {Spence}(2013)}]{Spence2013}%
  \BibitemOpen
  \bibfield  {author} {\bibinfo {author} {\bibfnamefont {J.}~\bibnamefont
  {Spence}},\ }\href@noop {} {\emph {\bibinfo {title} {High-{R}esolution
  {E}lectron {M}icroscopy}}}\ (\bibinfo  {publisher} {Oxford University
  Press},\ \bibinfo {year} {2013})\BibitemShut {NoStop}%
\bibitem [{\citenamefont {Zewail}\ and\ \citenamefont
  {Thomas}(2010)}]{Zewail2010a}%
  \BibitemOpen
  \bibfield  {author} {\bibinfo {author} {\bibfnamefont {A.~H.}\ \bibnamefont
  {Zewail}}\ and\ \bibinfo {author} {\bibfnamefont {J.~M.}\ \bibnamefont
  {Thomas}},\ }\href@noop {} {\emph {\bibinfo {title} {4D Electron Microscopy
  Imaging in Space and Time}}}\ (\bibinfo  {publisher} {Imperial College
  Press},\ \bibinfo {year} {2010})\BibitemShut {NoStop}%
\bibitem [{\citenamefont {King}\ \emph {et~al.}(2005)\citenamefont {King},
  \citenamefont {Campbell}, \citenamefont {Frank}, \citenamefont {Reed},
  \citenamefont {Schmerge}, \citenamefont {Siwick}, \citenamefont {Stuart},\
  and\ \citenamefont {Weber}}]{King2005}%
  \BibitemOpen
  \bibfield  {author} {\bibinfo {author} {\bibfnamefont {W.~E.}\ \bibnamefont
  {King}}, \bibinfo {author} {\bibfnamefont {G.~H.}\ \bibnamefont {Campbell}},
  \bibinfo {author} {\bibfnamefont {A.}~\bibnamefont {Frank}}, \bibinfo
  {author} {\bibfnamefont {B.}~\bibnamefont {Reed}}, \bibinfo {author}
  {\bibfnamefont {J.~F.}\ \bibnamefont {Schmerge}}, \bibinfo {author}
  {\bibfnamefont {B.~J.}\ \bibnamefont {Siwick}}, \bibinfo {author}
  {\bibfnamefont {B.~C.}\ \bibnamefont {Stuart}}, \ and\ \bibinfo {author}
  {\bibfnamefont {P.~M.}\ \bibnamefont {Weber}},\ }\href@noop {} {\bibfield
  {journal} {\bibinfo  {journal} {J.~Appl.~Phys.}\ }\textbf {\bibinfo {volume}
  {97}},\ \bibinfo {eid} {111101} (\bibinfo {year} {2005})}\BibitemShut
  {NoStop}%
\bibitem [{\citenamefont {Sciaini}\ and\ \citenamefont
  {Miller}(2011)}]{Sciaini2011}%
  \BibitemOpen
  \bibfield  {author} {\bibinfo {author} {\bibfnamefont {G.}~\bibnamefont
  {Sciaini}}\ and\ \bibinfo {author} {\bibfnamefont {R.~J.~D.}\ \bibnamefont
  {Miller}},\ }\href@noop {} {\bibfield  {journal} {\bibinfo  {journal}
  {Rep.~Progr.~Phys.}\ }\textbf {\bibinfo {volume} {74}},\ \bibinfo {pages}
  {096101} (\bibinfo {year} {2011})}\BibitemShut {NoStop}%
\bibitem [{\citenamefont {Baum}\ \emph {et~al.}(2007)\citenamefont {Baum},
  \citenamefont {Yang},\ and\ \citenamefont {Zewail}}]{Baum2007b}%
  \BibitemOpen
  \bibfield  {author} {\bibinfo {author} {\bibfnamefont {P.}~\bibnamefont
  {Baum}}, \bibinfo {author} {\bibfnamefont {D.-S.}\ \bibnamefont {Yang}}, \
  and\ \bibinfo {author} {\bibfnamefont {A.~H.}\ \bibnamefont {Zewail}},\
  }\href@noop {} {\bibfield  {journal} {\bibinfo  {journal} {Science}\ }\textbf
  {\bibinfo {volume} {318}},\ \bibinfo {pages} {788} (\bibinfo {year}
  {2007})}\BibitemShut {NoStop}%
\bibitem [{\citenamefont {Yurtsever}\ \emph {et~al.}(2012)\citenamefont
  {Yurtsever}, \citenamefont {van~der Veen},\ and\ \citenamefont
  {Zewail}}]{Yurtsever2012}%
  \BibitemOpen
  \bibfield  {author} {\bibinfo {author} {\bibfnamefont {A.}~\bibnamefont
  {Yurtsever}}, \bibinfo {author} {\bibfnamefont {R.~M.}\ \bibnamefont {van~der
  Veen}}, \ and\ \bibinfo {author} {\bibfnamefont {A.~H.}\ \bibnamefont
  {Zewail}},\ }\href@noop {} {\bibfield  {journal} {\bibinfo  {journal}
  {Science}\ }\textbf {\bibinfo {volume} {335}},\ \bibinfo {pages} {59}
  (\bibinfo {year} {2012})}\BibitemShut {NoStop}%
\bibitem [{\citenamefont {Erni}\ \emph {et~al.}(2009)\citenamefont {Erni},
  \citenamefont {Rossell}, \citenamefont {Kisielowski},\ and\ \citenamefont
  {Dahmen}}]{Erni2009}%
  \BibitemOpen
  \bibfield  {author} {\bibinfo {author} {\bibfnamefont {R.}~\bibnamefont
  {Erni}}, \bibinfo {author} {\bibfnamefont {M.~D.}\ \bibnamefont {Rossell}},
  \bibinfo {author} {\bibfnamefont {C.}~\bibnamefont {Kisielowski}}, \ and\
  \bibinfo {author} {\bibfnamefont {U.}~\bibnamefont {Dahmen}},\ }\href@noop {}
  {\bibfield  {journal} {\bibinfo  {journal} {Phys.~Rev.~Lett.}\ }\textbf
  {\bibinfo {volume} {102}},\ \bibinfo {pages} {096101} (\bibinfo {year}
  {2009})}\BibitemShut {NoStop}%
\bibitem [{\citenamefont {Crewe}\ \emph {et~al.}(1968)\citenamefont {Crewe},
  \citenamefont {Eggenberger}, \citenamefont {Wall},\ and\ \citenamefont
  {Welter}}]{Crewe1968}%
  \BibitemOpen
  \bibfield  {author} {\bibinfo {author} {\bibfnamefont {A.~V.}\ \bibnamefont
  {Crewe}}, \bibinfo {author} {\bibfnamefont {D.~N.}\ \bibnamefont
  {Eggenberger}}, \bibinfo {author} {\bibfnamefont {J.}~\bibnamefont {Wall}}, \
  and\ \bibinfo {author} {\bibfnamefont {L.~M.}\ \bibnamefont {Welter}},\
  }\href@noop {} {\bibfield  {journal} {\bibinfo  {journal} {Rev. Sci.
  Instrum.}\ }\textbf {\bibinfo {volume} {39}},\ \bibinfo {pages} {576}
  (\bibinfo {year} {1968})}\BibitemShut {NoStop}%
\bibitem [{\citenamefont {Lichte}\ and\ \citenamefont
  {Lehmann}(2008)}]{Lichte2008}%
  \BibitemOpen
  \bibfield  {author} {\bibinfo {author} {\bibfnamefont {H.}~\bibnamefont
  {Lichte}}\ and\ \bibinfo {author} {\bibfnamefont {M.}~\bibnamefont
  {Lehmann}},\ }\href@noop {} {\bibfield  {journal} {\bibinfo  {journal}
  {Rep.~Progr.~Phys.}\ }\textbf {\bibinfo {volume} {71}},\ \bibinfo {pages}
  {016102} (\bibinfo {year} {2008})}\BibitemShut {NoStop}%
\bibitem [{\citenamefont {Hasselbach}(2010)}]{Hasselbach2010}%
  \BibitemOpen
  \bibfield  {author} {\bibinfo {author} {\bibfnamefont {F.}~\bibnamefont
  {Hasselbach}},\ }\href@noop {} {\bibfield  {journal} {\bibinfo  {journal}
  {Rep.~Progr.~Phys.}\ }\textbf {\bibinfo {volume} {73}},\ \bibinfo {pages}
  {016101} (\bibinfo {year} {2010})}\BibitemShut {NoStop}%
\bibitem [{\citenamefont {Lougovski}\ and\ \citenamefont
  {Batelaan}(2011)}]{Lougovski2011}%
  \BibitemOpen
  \bibfield  {author} {\bibinfo {author} {\bibfnamefont {P.}~\bibnamefont
  {Lougovski}}\ and\ \bibinfo {author} {\bibfnamefont {H.}~\bibnamefont
  {Batelaan}},\ }\href {\doibase 10.1103/PhysRevA.84.023417} {\bibfield
  {journal} {\bibinfo  {journal} {Phys.~Rev.~A}\ }\textbf {\bibinfo {volume}
  {84}},\ \bibinfo {pages} {023417} (\bibinfo {year} {2011})}\BibitemShut
  {NoStop}%
\bibitem [{\citenamefont {Cho}\ \emph {et~al.}(2004)\citenamefont {Cho},
  \citenamefont {Ichimura}, \citenamefont {Shimizu},\ and\ \citenamefont
  {Oshima}}]{Cho2004}%
  \BibitemOpen
  \bibfield  {author} {\bibinfo {author} {\bibfnamefont {B.}~\bibnamefont
  {Cho}}, \bibinfo {author} {\bibfnamefont {T.}~\bibnamefont {Ichimura}},
  \bibinfo {author} {\bibfnamefont {R.}~\bibnamefont {Shimizu}}, \ and\
  \bibinfo {author} {\bibfnamefont {C.}~\bibnamefont {Oshima}},\ }\href@noop {}
  {\bibfield  {journal} {\bibinfo  {journal} {Phys.~Rev.~Lett.}\ }\textbf
  {\bibinfo {volume} {92}},\ \bibinfo {pages} {246103} (\bibinfo {year}
  {2004})}\BibitemShut {NoStop}%
\bibitem [{\citenamefont {Jensen}\ \emph {et~al.}(2010)\citenamefont {Jensen},
  \citenamefont {O'Shea}, \citenamefont {Feldman},\ and\ \citenamefont
  {Shaw}}]{Jensen2010}%
  \BibitemOpen
  \bibfield  {author} {\bibinfo {author} {\bibfnamefont {K.~L.}\ \bibnamefont
  {Jensen}}, \bibinfo {author} {\bibfnamefont {P.~G.}\ \bibnamefont {O'Shea}},
  \bibinfo {author} {\bibfnamefont {D.~W.}\ \bibnamefont {Feldman}}, \ and\
  \bibinfo {author} {\bibfnamefont {J.~L.}\ \bibnamefont {Shaw}},\ }\href@noop
  {} {\bibfield  {journal} {\bibinfo  {journal} {J.~Appl.~Phys.}\ }\textbf
  {\bibinfo {volume} {107}},\ \bibinfo {eid} {014903} (\bibinfo {year}
  {2010})}\BibitemShut {NoStop}%
\bibitem [{\citenamefont {Fowler}\ and\ \citenamefont
  {Nordheim}(1928)}]{Fowler1928}%
  \BibitemOpen
  \bibfield  {author} {\bibinfo {author} {\bibfnamefont {R.~H.}\ \bibnamefont
  {Fowler}}\ and\ \bibinfo {author} {\bibfnamefont {L.}~\bibnamefont
  {Nordheim}},\ }\href@noop {} {\bibfield  {journal} {\bibinfo  {journal}
  {Proc. R. Soc. London A}\ }\textbf {\bibinfo {volume} {119}},\ \bibinfo
  {pages} {173} (\bibinfo {year} {1928})}\BibitemShut {NoStop}%
\bibitem [{\citenamefont {Delone}\ and\ \citenamefont
  {Krainov}(1994)}]{Delone1994}%
  \BibitemOpen
  \bibfield  {author} {\bibinfo {author} {\bibfnamefont {N.~B.}\ \bibnamefont
  {Delone}}\ and\ \bibinfo {author} {\bibfnamefont {V.~P.}\ \bibnamefont
  {Krainov}},\ }\href@noop {} {\emph {\bibinfo {title} {Multiphoton Processes
  in Atoms}}}\ (\bibinfo  {publisher} {Springer},\ \bibinfo {address} {Berlin
  Heidelberg New York},\ \bibinfo {year} {1994})\BibitemShut {NoStop}%
\bibitem [{\citenamefont {Cook}\ \emph {et~al.}(2010)\citenamefont {Cook},
  \citenamefont {Verduin}, \citenamefont {Hagen},\ and\ \citenamefont
  {Kruit}}]{Cook2010}%
  \BibitemOpen
  \bibfield  {author} {\bibinfo {author} {\bibfnamefont {B.}~\bibnamefont
  {Cook}}, \bibinfo {author} {\bibfnamefont {T.}~\bibnamefont {Verduin}},
  \bibinfo {author} {\bibfnamefont {C.~W.}\ \bibnamefont {Hagen}}, \ and\
  \bibinfo {author} {\bibfnamefont {P.}~\bibnamefont {Kruit}},\ }\href@noop {}
  {\bibfield  {journal} {\bibinfo  {journal} {J.~Vac.~Sci.~Tech.~B}\ }\textbf
  {\bibinfo {volume} {28}},\ \bibinfo {pages} {C6C74} (\bibinfo {year}
  {2010})}\BibitemShut {NoStop}%
\bibitem [{\citenamefont {Beyer}\ and\ \citenamefont
  {G{\"o}lzh{\"a}user}(2010)}]{Beyer2010}%
  \BibitemOpen
  \bibfield  {author} {\bibinfo {author} {\bibfnamefont {A.}~\bibnamefont
  {Beyer}}\ and\ \bibinfo {author} {\bibfnamefont {A.}~\bibnamefont
  {G{\"o}lzh{\"a}user}},\ }\href {\doibase 10.1088/0953-8984/22/34/343001}
  {\bibfield  {journal} {\bibinfo  {journal} {J. of Physics: Condensed Matter}\
  }\textbf {\bibinfo {volume} {22}},\ \bibinfo {pages} {343001} (\bibinfo
  {year} {2010})}\BibitemShut {NoStop}%
\bibitem [{\citenamefont {Longchamp}\ \emph {et~al.}(2013)\citenamefont
  {Longchamp}, \citenamefont {Latychevskaia}, \citenamefont {Escher},\ and\
  \citenamefont {Fink}}]{Longchamp2013}%
  \BibitemOpen
  \bibfield  {author} {\bibinfo {author} {\bibfnamefont {J.-N.}\ \bibnamefont
  {Longchamp}}, \bibinfo {author} {\bibfnamefont {T.}~\bibnamefont
  {Latychevskaia}}, \bibinfo {author} {\bibfnamefont {C.}~\bibnamefont
  {Escher}}, \ and\ \bibinfo {author} {\bibfnamefont {H.~W.}\ \bibnamefont
  {Fink}},\ }\href@noop {} {\bibfield  {journal} {\bibinfo  {journal}
  {Phys.~Rev.~Lett.}\ }\textbf {\bibinfo {volume} {110}},\ \bibinfo {pages}
  {22501} (\bibinfo {year} {2013})}\BibitemShut {NoStop}%
\bibitem [{\citenamefont {Spence}\ \emph {et~al.}(1994)\citenamefont {Spence},
  \citenamefont {Qian},\ and\ \citenamefont {Silverman}}]{Spence1994}%
  \BibitemOpen
  \bibfield  {author} {\bibinfo {author} {\bibfnamefont {J.~C.~H.}\
  \bibnamefont {Spence}}, \bibinfo {author} {\bibfnamefont {W.}~\bibnamefont
  {Qian}}, \ and\ \bibinfo {author} {\bibfnamefont {M.~P.}\ \bibnamefont
  {Silverman}},\ }\href@noop {} {\bibfield  {journal} {\bibinfo  {journal}
  {J.~Vac.~Sci.~Tech.~A}\ }\textbf {\bibinfo {volume} {12}},\ \bibinfo {pages}
  {542} (\bibinfo {year} {1994})}\BibitemShut {NoStop}%
\bibitem [{\citenamefont {Pozzi}(1987)}]{Pozzi1987}%
  \BibitemOpen
  \bibfield  {author} {\bibinfo {author} {\bibfnamefont {G.}~\bibnamefont
  {Pozzi}},\ }\href@noop {} {\bibfield  {journal} {\bibinfo  {journal} {Optik}\
  }\textbf {\bibinfo {volume} {77}},\ \bibinfo {pages} {69} (\bibinfo {year}
  {1987})}\BibitemShut {NoStop}%
\bibitem [{\citenamefont {Barwick}\ \emph {et~al.}(2008)\citenamefont
  {Barwick}, \citenamefont {Park}, \citenamefont {Kwon}, \citenamefont
  {Baskin},\ and\ \citenamefont {Zewail}}]{Barwick2008}%
  \BibitemOpen
  \bibfield  {author} {\bibinfo {author} {\bibfnamefont {B.}~\bibnamefont
  {Barwick}}, \bibinfo {author} {\bibfnamefont {H.~S.}\ \bibnamefont {Park}},
  \bibinfo {author} {\bibfnamefont {O.-H.}\ \bibnamefont {Kwon}}, \bibinfo
  {author} {\bibfnamefont {J.~S.}\ \bibnamefont {Baskin}}, \ and\ \bibinfo
  {author} {\bibfnamefont {A.~H.}\ \bibnamefont {Zewail}},\ }\href {\doibase
  10.1126/science.1164000} {\bibfield  {journal} {\bibinfo  {journal}
  {Science}\ }\textbf {\bibinfo {volume} {322}},\ \bibinfo {pages} {1227}
  (\bibinfo {year} {2008})}\BibitemShut {NoStop}%
\bibitem [{Not()}]{NoteCommercialLasertriggeredSEM}%
  \BibitemOpen
  \href@noop {} {}\bibinfo {note} {Despite the state of development, a
  femtosecond laser-triggered electron microscope is commercially available
  from a large electron microscope manufacturer, indicating the expected market
  volume. See www.fei.com/tecnai-femto/}\BibitemShut {NoStop}%
\bibitem [{\citenamefont {Yang}\ \emph {et~al.}(2010)\citenamefont {Yang},
  \citenamefont {Mohammed},\ and\ \citenamefont {Zewail}}]{Yang2010}%
  \BibitemOpen
  \bibfield  {author} {\bibinfo {author} {\bibfnamefont {D.-S.}\ \bibnamefont
  {Yang}}, \bibinfo {author} {\bibfnamefont {O.~F.}\ \bibnamefont {Mohammed}},
  \ and\ \bibinfo {author} {\bibfnamefont {A.~H.}\ \bibnamefont {Zewail}},\
  }\href@noop {} {\bibfield  {journal} {\bibinfo  {journal}
  {Proc.~Nat.~Acad.~Sci.~USA}\ }\textbf {\bibinfo {volume} {107}},\ \bibinfo
  {pages} {14993} (\bibinfo {year} {2010})}\BibitemShut {NoStop}%
\bibitem [{\citenamefont {Kirchner}\ \emph {et~al.}(2014)\citenamefont
  {Kirchner}, \citenamefont {Gliserin}, \citenamefont {Krausz},\ and\
  \citenamefont {Baum}}]{Kirchner2014}%
  \BibitemOpen
  \bibfield  {author} {\bibinfo {author} {\bibfnamefont {F.~O.}\ \bibnamefont
  {Kirchner}}, \bibinfo {author} {\bibfnamefont {A.}~\bibnamefont {Gliserin}},
  \bibinfo {author} {\bibfnamefont {F.}~\bibnamefont {Krausz}}, \ and\ \bibinfo
  {author} {\bibfnamefont {P.}~\bibnamefont {Baum}},\ }\href@noop {} {\bibfield
   {journal} {\bibinfo  {journal} {Nat. Photon}\ }\textbf {\bibinfo {volume}
  {8}},\ \bibinfo {pages} {52} (\bibinfo {year} {2014})}\BibitemShut {NoStop}%
\bibitem [{\citenamefont {Quinonez}\ \emph {et~al.}(2013)\citenamefont
  {Quinonez}, \citenamefont {Handali},\ and\ \citenamefont
  {Barwick}}]{Quinonez2013}%
  \BibitemOpen
  \bibfield  {author} {\bibinfo {author} {\bibfnamefont {E.}~\bibnamefont
  {Quinonez}}, \bibinfo {author} {\bibfnamefont {J.}~\bibnamefont {Handali}}, \
  and\ \bibinfo {author} {\bibfnamefont {B.}~\bibnamefont {Barwick}},\
  }\href@noop {} {\bibfield  {journal} {\bibinfo  {journal} {Rev.~Sci.~Instr.}\
  }\textbf {\bibinfo {volume} {84}},\ \bibinfo {eid} {103710} (\bibinfo {year}
  {2013})}\BibitemShut {NoStop}%
\bibitem [{\citenamefont {Gulde}\ \emph {et~al.}(2014)\citenamefont {Gulde},
  \citenamefont {Schweda}, \citenamefont {Storeck}, \citenamefont {Maiti},
  \citenamefont {Yu}, \citenamefont {Wodtke}, \citenamefont {Sch{\"a}fer},\
  and\ \citenamefont {Ropers}}]{Gulde2014}%
  \BibitemOpen
  \bibfield  {author} {\bibinfo {author} {\bibfnamefont {M.}~\bibnamefont
  {Gulde}}, \bibinfo {author} {\bibfnamefont {S.}~\bibnamefont {Schweda}},
  \bibinfo {author} {\bibfnamefont {G.}~\bibnamefont {Storeck}}, \bibinfo
  {author} {\bibfnamefont {M.}~\bibnamefont {Maiti}}, \bibinfo {author}
  {\bibfnamefont {H.~K.}\ \bibnamefont {Yu}}, \bibinfo {author} {\bibfnamefont
  {A.~M.}\ \bibnamefont {Wodtke}}, \bibinfo {author} {\bibfnamefont
  {S.}~\bibnamefont {Sch{\"a}fer}}, \ and\ \bibinfo {author} {\bibfnamefont
  {C.}~\bibnamefont {Ropers}},\ }\href@noop {} {\bibfield  {journal} {\bibinfo
  {journal} {Science}\ }\textbf {\bibinfo {volume} {345}},\ \bibinfo {pages}
  {200} (\bibinfo {year} {2014})}\BibitemShut {NoStop}%
\bibitem [{\citenamefont {M{\"u}ller}\ \emph {et~al.}(2014)\citenamefont
  {M{\"u}ller}, \citenamefont {Paarmann},\ and\ \citenamefont
  {Ernstorfer}}]{Muller2014}%
  \BibitemOpen
  \bibfield  {author} {\bibinfo {author} {\bibfnamefont {M.}~\bibnamefont
  {M{\"u}ller}}, \bibinfo {author} {\bibfnamefont {A.}~\bibnamefont
  {Paarmann}}, \ and\ \bibinfo {author} {\bibfnamefont {R.}~\bibnamefont
  {Ernstorfer}},\ }\href@noop {} {\bibfield  {journal} {\bibinfo  {journal}
  {ArXiv e-prints}\ } (\bibinfo {year} {2014})},\ \Eprint
  {http://arxiv.org/abs/1405.4992} {arXiv:1405.4992} \BibitemShut {NoStop}%
\bibitem [{\citenamefont {Hommelhoff}\ \emph
  {et~al.}(2006{\natexlab{a}})\citenamefont {Hommelhoff}, \citenamefont
  {Sortais}, \citenamefont {Aghajani-Talesh},\ and\ \citenamefont
  {Kasevich}}]{Hommelhoff2006}%
  \BibitemOpen
  \bibfield  {author} {\bibinfo {author} {\bibfnamefont {P.}~\bibnamefont
  {Hommelhoff}}, \bibinfo {author} {\bibfnamefont {Y.}~\bibnamefont {Sortais}},
  \bibinfo {author} {\bibfnamefont {A.}~\bibnamefont {Aghajani-Talesh}}, \ and\
  \bibinfo {author} {\bibfnamefont {M.~A.}\ \bibnamefont {Kasevich}},\ }\href
  {\doibase 10.1103/PhysRevLett.96.077401} {\bibfield  {journal} {\bibinfo
  {journal} {Phys.~Rev.~Lett.}\ }\textbf {\bibinfo {volume} {96}},\ \bibinfo
  {pages} {077401} (\bibinfo {year} {2006}{\natexlab{a}})}\BibitemShut
  {NoStop}%
\bibitem [{\citenamefont {Hommelhoff}\ \emph
  {et~al.}(2006{\natexlab{b}})\citenamefont {Hommelhoff}, \citenamefont
  {Kealhofer},\ and\ \citenamefont {Kasevich}}]{Hommelhoff2006a}%
  \BibitemOpen
  \bibfield  {author} {\bibinfo {author} {\bibfnamefont {P.}~\bibnamefont
  {Hommelhoff}}, \bibinfo {author} {\bibfnamefont {C.}~\bibnamefont
  {Kealhofer}}, \ and\ \bibinfo {author} {\bibfnamefont {M.~A.}\ \bibnamefont
  {Kasevich}},\ }\href@noop {} {\bibfield  {journal} {\bibinfo  {journal}
  {Phys.~Rev.~Lett.}\ }\textbf {\bibinfo {volume} {97}},\ \bibinfo {eid}
  {247402} (\bibinfo {year} {2006}{\natexlab{b}})}\BibitemShut {NoStop}%
\bibitem [{\citenamefont {Ropers}\ \emph {et~al.}(2007)\citenamefont {Ropers},
  \citenamefont {Solli}, \citenamefont {Schulz}, \citenamefont {Lienau},\ and\
  \citenamefont {Elsaesser}}]{Ropers2007}%
  \BibitemOpen
  \bibfield  {author} {\bibinfo {author} {\bibfnamefont {C.}~\bibnamefont
  {Ropers}}, \bibinfo {author} {\bibfnamefont {D.~R.}\ \bibnamefont {Solli}},
  \bibinfo {author} {\bibfnamefont {C.~P.}\ \bibnamefont {Schulz}}, \bibinfo
  {author} {\bibfnamefont {C.}~\bibnamefont {Lienau}}, \ and\ \bibinfo {author}
  {\bibfnamefont {T.}~\bibnamefont {Elsaesser}},\ }\href {\doibase
  10.1103/PhysRevLett.98.043907} {\bibfield  {journal} {\bibinfo  {journal}
  {Phys.~Rev.~Lett.}\ }\textbf {\bibinfo {volume} {98}},\ \bibinfo {pages}
  {043907} (\bibinfo {year} {2007})}\BibitemShut {NoStop}%
\bibitem [{\citenamefont {Peralta}\ \emph {et~al.}(2013)\citenamefont
  {Peralta}, \citenamefont {Soong}, \citenamefont {England}, \citenamefont
  {Wu}, \citenamefont {Montazeri}, \citenamefont {McGuinness}, \citenamefont
  {McNeur}, \citenamefont {Leedle}, \citenamefont {Walz}, \citenamefont
  {Sozer}, \citenamefont {Cowan}, \citenamefont {Schwartz}, \citenamefont
  {Travish},\ and\ \citenamefont {Byer}}]{Peralta2013}%
  \BibitemOpen
  \bibfield  {author} {\bibinfo {author} {\bibfnamefont {E.}~\bibnamefont
  {Peralta}}, \bibinfo {author} {\bibfnamefont {K.}~\bibnamefont {Soong}},
  \bibinfo {author} {\bibfnamefont {E.~R.}\ \bibnamefont {England},
  \bibfnamefont {R.~J.and~Colby}}, \bibinfo {author} {\bibfnamefont
  {Z.}~\bibnamefont {Wu}}, \bibinfo {author} {\bibfnamefont {B.}~\bibnamefont
  {Montazeri}}, \bibinfo {author} {\bibfnamefont {C.}~\bibnamefont
  {McGuinness}}, \bibinfo {author} {\bibfnamefont {J.}~\bibnamefont {McNeur}},
  \bibinfo {author} {\bibfnamefont {K.~J.}\ \bibnamefont {Leedle}}, \bibinfo
  {author} {\bibfnamefont {D.}~\bibnamefont {Walz}}, \bibinfo {author}
  {\bibfnamefont {E.}~\bibnamefont {Sozer}}, \bibinfo {author} {\bibfnamefont
  {B.}~\bibnamefont {Cowan}}, \bibinfo {author} {\bibfnamefont
  {B.}~\bibnamefont {Schwartz}}, \bibinfo {author} {\bibfnamefont
  {G.}~\bibnamefont {Travish}}, \ and\ \bibinfo {author} {\bibfnamefont
  {R.~L.}\ \bibnamefont {Byer}},\ }\href@noop {} {\bibfield  {journal}
  {\bibinfo  {journal} {Nature}\ }\textbf {\bibinfo {volume} {503}},\ \bibinfo
  {pages} {91} (\bibinfo {year} {2013})}\BibitemShut {NoStop}%
\bibitem [{\citenamefont {Breuer}\ and\ \citenamefont
  {Hommelhoff}(2013)}]{Breuer2013}%
  \BibitemOpen
  \bibfield  {author} {\bibinfo {author} {\bibfnamefont {J.}~\bibnamefont
  {Breuer}}\ and\ \bibinfo {author} {\bibfnamefont {P.}~\bibnamefont
  {Hommelhoff}},\ }\href {\doibase 10.1103/PhysRevLett.111.134803} {\bibfield
  {journal} {\bibinfo  {journal} {Phys.~Rev.~Lett.}\ }\textbf {\bibinfo
  {volume} {111}},\ \bibinfo {pages} {134803} (\bibinfo {year}
  {2013})}\BibitemShut {NoStop}%
\end{thebibliography}%


\begin{thebibliography}{4}%
\makeatletter
\providecommand \@ifxundefined [1]{%
 \@ifx{#1\undefined}
}%
\providecommand \@ifnum [1]{%
 \ifnum #1\expandafter \@firstoftwo
 \else \expandafter \@secondoftwo
 \fi
}%
\providecommand \@ifx [1]{%
 \ifx #1\expandafter \@firstoftwo
 \else \expandafter \@secondoftwo
 \fi
}%
\providecommand \natexlab [1]{#1}%
\providecommand \enquote  [1]{``#1''}%
\providecommand \bibnamefont  [1]{#1}%
\providecommand \bibfnamefont [1]{#1}%
\providecommand \citenamefont [1]{#1}%
\providecommand \href@noop [0]{\@secondoftwo}%
\providecommand \href [0]{\begingroup \@sanitize@url \@href}%
\providecommand \@href[1]{\@@startlink{#1}\@@href}%
\providecommand \@@href[1]{\endgroup#1\@@endlink}%
\providecommand \@sanitize@url [0]{\catcode `\\12\catcode `\$12\catcode
  `\&12\catcode `\#12\catcode `\^12\catcode `\_12\catcode `\%12\relax}%
\providecommand \@@startlink[1]{}%
\providecommand \@@endlink[0]{}%
\providecommand \url  [0]{\begingroup\@sanitize@url \@url }%
\providecommand \@url [1]{\endgroup\@href {#1}{\urlprefix }}%
\providecommand \urlprefix  [0]{URL }%
\providecommand \Eprint [0]{\href }%
\providecommand \doibase [0]{http://dx.doi.org/}%
\providecommand \selectlanguage [0]{\@gobble}%
\providecommand \bibinfo  [0]{\@secondoftwo}%
\providecommand \bibfield  [0]{\@secondoftwo}%
\providecommand \translation [1]{[#1]}%
\providecommand \BibitemOpen [0]{}%
\providecommand \bibitemStop [0]{}%
\providecommand \bibitemNoStop [0]{.\EOS\space}%
\providecommand \EOS [0]{\spacefactor3000\relax}%
\providecommand \BibitemShut  [1]{\csname bibitem#1\endcsname}%
\let\auto@bib@innerbib\@empty
\bibitem [{\citenamefont {Yang}\ \emph {et~al.}(2010)\citenamefont {Yang},
  \citenamefont {Mohammed},\ and\ \citenamefont {Zewail}}]{Yang2010}%
  \BibitemOpen
  \bibfield  {author} {\bibinfo {author} {\bibfnamefont {D.-S.}\ \bibnamefont
  {Yang}}, \bibinfo {author} {\bibfnamefont {O.~F.}\ \bibnamefont {Mohammed}},
  \ and\ \bibinfo {author} {\bibfnamefont {A.~H.}\ \bibnamefont {Zewail}},\
  }\href@noop {} {\bibfield  {journal} {\bibinfo  {journal}
  {Proc.~Nat.~Acad.~Sci.~USA}\ }\textbf {\bibinfo {volume} {107}},\ \bibinfo
  {pages} {14993} (\bibinfo {year} {2010})}\BibitemShut {NoStop}%
\bibitem [{\citenamefont {Hofmann}\ \emph {et~al.}(2013)\citenamefont
  {Hofmann}, \citenamefont {Gl{\"u}ckert}, \citenamefont {No{\'e}},
  \citenamefont {Bourjau}, \citenamefont {Dehmel},\ and\ \citenamefont
  {H{\"o}gele}}]{Hofmann2013}%
  \BibitemOpen
  \bibfield  {author} {\bibinfo {author} {\bibfnamefont {M.~S.}\ \bibnamefont
  {Hofmann}}, \bibinfo {author} {\bibfnamefont {J.~T.}\ \bibnamefont
  {Gl{\"u}ckert}}, \bibinfo {author} {\bibfnamefont {J.}~\bibnamefont
  {No{\'e}}}, \bibinfo {author} {\bibfnamefont {C.}~\bibnamefont {Bourjau}},
  \bibinfo {author} {\bibfnamefont {R.}~\bibnamefont {Dehmel}}, \ and\ \bibinfo
  {author} {\bibfnamefont {A.}~\bibnamefont {H{\"o}gele}},\ }\href@noop {}
  {\bibfield  {journal} {\bibinfo  {journal} {Nature Nanotechnology}\ }\textbf
  {\bibinfo {volume} {8}},\ \bibinfo {pages} {502} (\bibinfo {year}
  {2013})}\BibitemShut {NoStop}%
\bibitem [{\citenamefont {Hwang}\ \emph {et~al.}(2013)\citenamefont {Hwang},
  \citenamefont {Chang}, \citenamefont {Lu}, \citenamefont {Liu}, \citenamefont
  {Chang}, \citenamefont {Lee}, \citenamefont {Jeng}, \citenamefont {Kuo},
  \citenamefont {Lin}, \citenamefont {Chang},\ and\ \citenamefont
  {Tsong}}]{Hwang2013}%
  \BibitemOpen
  \bibfield  {author} {\bibinfo {author} {\bibfnamefont {I.-S.}\ \bibnamefont
  {Hwang}}, \bibinfo {author} {\bibfnamefont {C.-C.}\ \bibnamefont {Chang}},
  \bibinfo {author} {\bibfnamefont {C.-H.}\ \bibnamefont {Lu}}, \bibinfo
  {author} {\bibfnamefont {S.-C.}\ \bibnamefont {Liu}}, \bibinfo {author}
  {\bibfnamefont {Y.-C.}\ \bibnamefont {Chang}}, \bibinfo {author}
  {\bibfnamefont {T.-K.}\ \bibnamefont {Lee}}, \bibinfo {author} {\bibfnamefont
  {H.-T.}\ \bibnamefont {Jeng}}, \bibinfo {author} {\bibfnamefont {H.-S.}\
  \bibnamefont {Kuo}}, \bibinfo {author} {\bibfnamefont {C.-Y.}\ \bibnamefont
  {Lin}}, \bibinfo {author} {\bibfnamefont {C.-S.}\ \bibnamefont {Chang}}, \
  and\ \bibinfo {author} {\bibfnamefont {T.~T.}\ \bibnamefont {Tsong}},\
  }\href@noop {} {\bibfield  {journal} {\bibinfo  {journal} {New~J.~Phys.}\
  }\textbf {\bibinfo {volume} {15}},\ \bibinfo {pages} {043015} (\bibinfo
  {year} {2013})}\BibitemShut {NoStop}%
\bibitem [{\citenamefont {M{\"o}llenstedt}\ and\ \citenamefont
  {D{\"u}ker}(1956)}]{Mollenstedt1956}%
  \BibitemOpen
  \bibfield  {author} {\bibinfo {author} {\bibfnamefont {G.}~\bibnamefont
  {M{\"o}llenstedt}}\ and\ \bibinfo {author} {\bibfnamefont {H.}~\bibnamefont
  {D{\"u}ker}},\ }\href {\doibase 10.1007/BF01326780} {\bibfield  {journal}
  {\bibinfo  {journal} {Z.~Phys.}\ }\textbf {\bibinfo {volume} {145}},\
  \bibinfo {pages} {377} (\bibinfo {year} {1956})}\BibitemShut {NoStop}%
\end{thebibliography}%
\end{document}


\title{Supplemental Material}

\author{Dominik Ehberger}
\affiliation{Friedrich Alexander University Erlangen-Nuremberg, Department of Physics, Staudtstr. 1, D-91508 Erlangen, Germany, EU}
\affiliation{Max Planck Institute for Quantum Optics, Hans-Kopfermann-Str. 1, D-85748 Garching/Munich, Germany, EU}

\author{Jakob Hammer}
\affiliation{Friedrich Alexander University Erlangen-Nuremberg, Department of Physics, Staudtstr. 1, D-91508 Erlangen, Germany, EU}
\affiliation{Max Planck Institute for Quantum Optics, Hans-Kopfermann-Str. 1, D-85748 Garching/Munich, Germany, EU}

\author{Max Eisele}
\affiliation{Max Planck Institute for Quantum Optics, Hans-Kopfermann-Str. 1, D-85748 Garching/Munich, Germany, EU}

\author{Michael Kr{\"u}ger}
\affiliation{Friedrich Alexander University Erlangen-Nuremberg, Department of Physics, Staudtstr. 1, D-91508 Erlangen, Germany, EU}
\affiliation{Max Planck Institute for Quantum Optics, Hans-Kopfermann-Str. 1, D-85748 Garching/Munich, Germany, EU}

\author{Jonathan Noe}
\affiliation{Fakult{\"a}t f{\"u}r Physik and Center for NanoScience (CeNS), Ludwig-Maximilians-Universit{\"a}t M{\"u}nchen, Geschwister-Scholl-Platz 1, 80539 M{\"u}nchen, Germany, EU}

\author{Alexander H{\"o}gele}
\affiliation{Fakult{\"a}t f{\"u}r Physik and Center for NanoScience (CeNS), Ludwig-Maximilians-Universit{\"a}t M{\"u}nchen, Geschwister-Scholl-Platz 1, 80539 M{\"u}nchen, Germany, EU}

\author{Peter Hommelhoff}
\affiliation{Friedrich Alexander University Erlangen-Nuremberg, Department of Physics, Staudtstr. 1, D-91508 Erlangen, Germany, EU}
\affiliation{Max Planck Institute for Quantum Optics, Hans-Kopfermann-Str. 1, D-85748 Garching/Munich, Germany, EU}
\affiliation{Max Planck Institute for the Science of Light, G\"unther-Scharowsky-Str. 1/ Bldg. 24, D-91508 Erlangen, Germany, EU}
\date{\today}

 
\setcounter{page}{5}

\section*{Supplementary Material}

\section{Experimental setup}
\label{sec: setup}
\begin{figure}
\includegraphics[width=86mm]{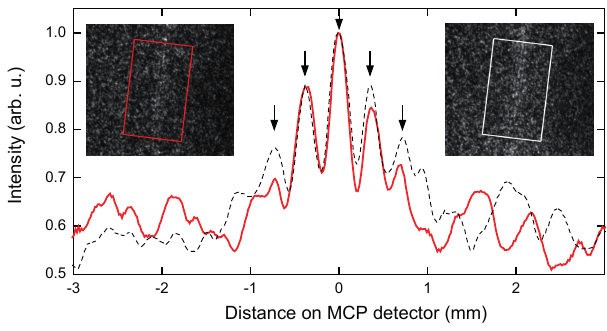}
\caption{Integrated line profiles of a single interference image of carbon nanotubes  with grin-lens focusing of the pulsed picosecond pulses (red solid line, left inset) and DC-field emission source operated at comparable current (black dashed line, right inset). Outermost fringes are not discernible because of mechanical vibration during the comparably long image acquisition time ($1.28\,$s) necessary due to the low laser intensity incident on the tip.}
\label{fig:lowpow}
\end{figure}

The whole experimental setup is mounted in a ultra-high vacuum chamber at a base pressure of $<7\times 10^{-8}\,$Pa. The chamber is mounted on a small ($\sim$1\,m$^2$) optical bench mounted on standard air cushion supports in a quiet lab. Tip, nano-positioners and MCP detector are housed in a box of \textmu -metal inside the chamber for magnetic shielding. Laser light was guided to the experiment with the help of a glass fiber. Focusing on the tip was achieved with the help of a gradient index (grin) lens (Grintech), achieving a focal spot of $4.7\,\mu$m $1/e^2$-radius. Optionally, a tighter focus of  $0.86\,\mu$m $1/e^2$-radius was achieved with a home-built assembly of an air-spaced achromatic doublet (\O~25.4\,mm, focal length 60\,mm) and an asphere (\O~20\,mm, 0.5\,numerical aperture). The laser pulses are derived from a Ti:sapph oscillator, frequency-doubled in a BiBO-crystal to 395\,nm and coupled into a polarization maintaining single-mode glass fibre. The fibre is inserted to the UHV chamber via a homebuilt fibre-vacuum feedthrough. In pulsed operation (calculated pulse duration of $21\,$ps at the fiber exit due to chromatic dispersion in the fibre), $5\,$mW of blue light are incident on the tip. An exemplary integrated line profile from an interference pattern obtained with this source and the grin-lens focusing is shown in Fig.~\ref{fig:lowpow} (red solid line) and for comparison with a DC-field emitter source at comparable current (black dashed line). The maximum attainable time averaged current in this configuration is 15\,fA, which corresponds to 0.03\,electrons per pulse.

The identical one-photon emission process is driven with the output of a blue (405\,nm) cw-laser diode, alternatively coupled into the fibre.
In order to minimize thermal drifts of the CNT  $\sim 5$\,mW of laser power were incident on the tip in this mode. However, together with the tighter focusing assembly this allows for increasing the photoemitted electron current by more than one order of magnitude. Interference patterns shown in Figs.~2 and~\ref{fig:lowmag} were obtained in this configuration.
At power levels up to $\sim 10\,$mW we did not see any tip instabilities, let alone any tip damage due to the laser beam. However, tips are easily damaged by crashing into the CNT sample.

The whole setup provides high stability and the alignment inside the chamber is fairly simple. For this reasons, we regard our fibre-coupled laser pulse delivery system as a prototype for laser-triggered microscopy columns, requiring high stability in the laser-tip alignment over typical acquisition times on the order of minutes~\cite{Yang2010}. Laser pulse durations on the tip on the order of 100\,fs can be achieved with standard dispersion compensation techniques and appropriate fibers.

\section{Sample preparation}
The substrate for supporting the CNTs is made of a $200\,$nm thin film of silicon nitride supported by a rigid silicon frame (DuraSiN, DTM-25231). Arrays of holes with a diameter of $2\,$\textmu m are arranged in a rectangular pattern with a pitch of $12\,$\textmu m. CNTs are grown on the substrate and over its holes employing a chemical vapour deposition process as described in detail in the supplement material of~\cite{Hofmann2013}. After examination of the sample in a scanning electron microscope it is coated with a $5\,$nm layer of gold on both sides using a $5\,$nm layer of titanium as adhesion promoter for the sample to act as a (grounded) counter electrode in the experiment.

\section{Image acquisition}
Electron interference images shown are comprised of 50 (Fig.~\ref{fig:lowmag}) up to 200 (Fig.~2a,b) individual images, each one recorded with an exposure times of $500\,$ms (Fig.~\ref{fig:lowmag}), $200\,$ms and $21\,$ms (Fig.~2a,b, respectively), with a standard 8-bit dynamic range CCD camera and a 35\,mm objective from outside of the vacuum chamber. Individual images are superimposed. Images may be shifted against each other in order to cancel effects of slow linear drifts. These are most likely induced by heating of the sample with the cw-laser beam and slow cooling during image acquisition with the DC field emission source (opposite signs of the drifts). The applied shift from image to image is found by fitting the central fringe of the pattern with a Gaussian function.

\section{Tip preparation}
The tips used in the experiment are electrochemically etched from a monocrystalline tungsten wire oriented along the [310]-direction. Typical tip radii obtained with this method are in the range of $5\dots20\,$nm. Field ion microscopy is employed for deducing the approximate radius of the tip and field evaporation cleaning from adsorbates. Furthermore, tips are also cleaned in situ by annealing prior to the conduction of the experiment.

\section{Mechanism of photoemission}

In order to obtain a large photocurrent the photoemission is driven closely below the threshold of DC field emission at $\sim 2\,$GV/m (Fig. 1a). In this case the DC current is still negligible ($<1\,$\%), but the photo-emitted current is maximized. At $1.4\dots 2\,$GV/m, the effective barrier height is reduced by the Schottky effect to $2.7\dots2.9\,$eV for the lowest workfunction plane ([310]).

We identify a one-photon emission process by measuring the emitted electron current as a function of the average laser power, which shows a linear dependence. Furthermore, the contribution of heating to the emission process is found to be negligible as the emitted current is maximized for light polarized parallel to the tip axis and minimized for light polarized perpendicular to it, in pulsed as well as in cw illumination mode.

In DC field emission electrons are predominantly emitted from the [310] crystallographic plane, which has the lowest workfunction. With $E_\mathrm{ph}=3.1\,$eV and DC-fields closely below the field emission threshold electrons from crystal planes with workfunctions up to $4.8\,$eV are photoemitted over the Schottky-lowered barrier (Fig. 1a). In the experiment this manifests itself by the higher divergence of the photo-emitted electron beam of $10^\circ$ (half opening angle) compared to $6^\circ$ in DC-field emission.

\section{Electron energy spectra analysis}
\label{sec:spec}
Energy spectra of the electrons photoemitted from the tip have been measured with a retarding field energy analyzer. Here as well, a frequency doubled Ti:sapph oscillator is employed, however at a slightly lower central photon energy of $E_\mathrm{ph}=2.8\,$eV compared to 3.1\,eV used for the interference measurements. The photon energy bandwidth of 0.06\,eV ($1/e^2$ value of the intensity) in both cases is about one order of magnitude smaller than the measured electron energy bandwidth. Hence, its contribution to the electron energy width can be neglected.

Primarily, the width of the electron spectra depends on the excess energy $\Delta EE$ of the emitted electrons.
It is given by
\begin{equation}
\Delta EE = E_\mathrm{ph}-\left(\phi_\mathrm{w}-\sqrt{\frac{e^3F}{4\pi\epsilon_0}}\right).
\end{equation}
The term in brackets is the effective barrier height, meaning the workfunction $\phi_\mathrm{w}$ reduced due to the Schottky effect with the elementary charge $e$, DC-field strength $F$ and the vacuum permittivity $\epsilon_0$.

For the effective barrier height of $2.8\pm0.1\,$eV in the interference measurements, we find for the respective excess energy in the spectral measurements an energy bandwidth of $0.51\pm0.06\,$eV (FWHM).

\section{Interference fringe analysis}

\begin{figure}
\includegraphics[width=86mm]{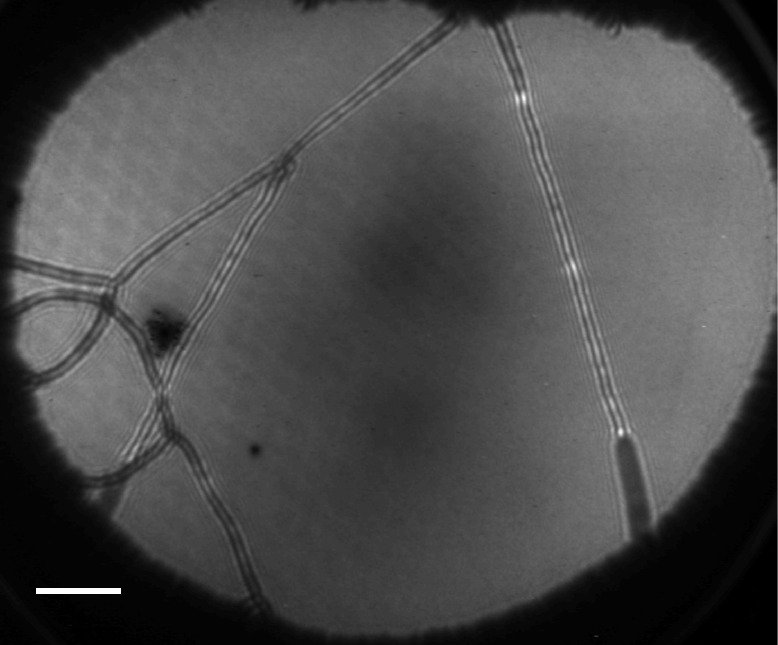}
\caption{Laser-triggered point projection image of carbon nanotubes spanning a hole of the carrier substrate at a tip-sample distance $a\approx5\,$\textmu m. A this magnification ($\sim 2\times 10^{4}$) biprism contributions (see text) are less pronounced. The scale bar represents 4\,mm on the MCP detector screen. The black spots on the left side of the image are due to defects on the detector.}
\label{fig:lowmag}
\end{figure}

\subsection{Interpretation of obtained projection images}

\begin{figure}
\includegraphics[width=86mm]{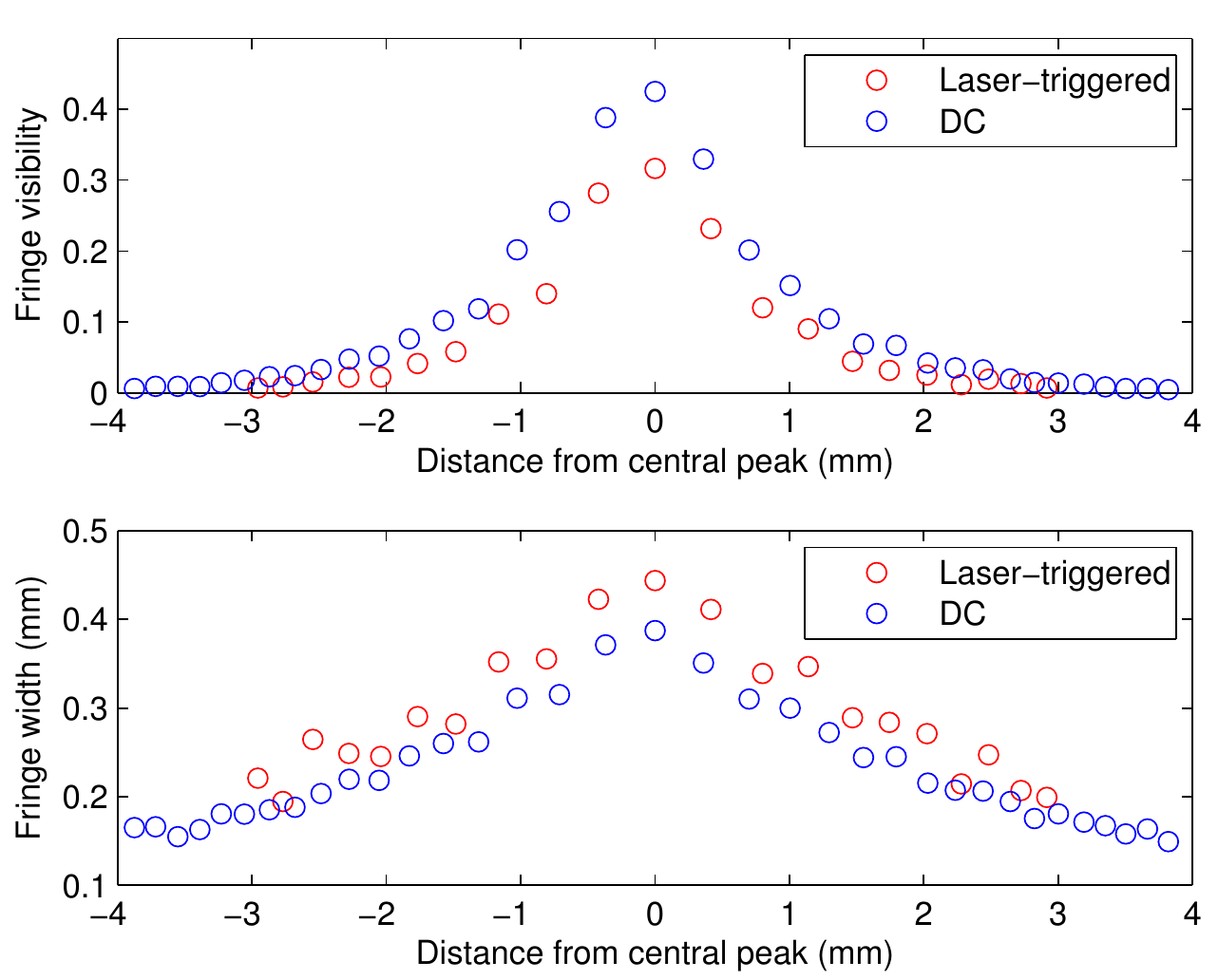}
\caption{Evaluation of interference patterns of Fig.~2c,d: fringe visibility and fringe width $w$ as function of distance from central peak. Higher visibility in the center in DC field emission is due to the slightly smaller effective source and smaller energy spread of the electron beam. The larger $w$ (defined as the distance between the two neighbouring minima) with the laser-triggered source in the center of the pattern reflects the larger wavelength of the  photo-emitted electrons. Further away from the center the difference becomes less severe because $w$ in Fresnel diffraction scales with $\sqrt{\lambda_\mathrm{dB}}$ as opposed to the biprism interference regime where $w$ is linear in $\lambda_\mathrm{dB}$.}
\label{fringespacing}
\end{figure}

The observable interference patterns can be distinguished into two regimes~\cite{Hwang2013}: (i) Holographic regime. For large tip-sample distances (low magnification) the projected image represents an in-line hologram of the CNT. The incident electron wave is scattered off the sample and interferes with the unscattered part of the wave, which gives rise to a Fresnel-like pattern with one broad fringe in its center (Fig.~\ref{fig:lowmag}). (ii) Biprism regime. With increasing magnification the static electric fields around the CNT start to play a major role. These lead to the deflection and subsequent overlapping in the detector plane of the two parts of the electron beam, which is split by the CNT. A cosinusoidal interference pattern is observed, equivalent to those obtained with an electrostatic biprism~\cite{Mollenstedt1956}. For intermediate magnifications regimes (i) and (ii) cannot be distinguished precisely any more, hence the patterns we observe in Fig.~2 show Fresnel and biprism contributions, which is evident from the three fringes in the center of comparable width and decreasing fringe spacing with increasing distance from the central peak (Fig.~\ref{fringespacing}).

\subsection{Interference fringe visibility}

The Michelson-visibility $V$ of interference fringes is defined as:
\begin{equation}
V=\frac{I_\mathrm{max}-I_\mathrm{min}}{I_\mathrm{max}+I_\mathrm{min}}
\end{equation}
with $I_\mathrm{max}$ ($I_\mathrm{max}$) being the maximum (minimum) intensities of neighbouring fringes. It directly reflects the coherence properties of the electron source. A monochromatic point source would yield a biprism interference pattern (see above) of $V=1$. One reason for the observed reduced visibility of the central fringe in laser-triggered mode 
with respect to DC-field emission mode 
by about $24\,$\% (Figs.~2 and~\ref{fringespacing}) is the larger effective source radius. Furthermore, the visibility decreases due to the larger energy bandwidth and therefore reduced longitudinal coherence of the emitted electron beam (see above).

\bibliography{scibib}